# Room temperature formation of high-mobility two-dimensional electron gases at crystalline complex oxide interfaces

Y. Z. Chen,[1]* N. Bovet,[2] T. Kasama,[3] W. W. Gao,[4] S. Yazdi,[3] C. Ma,[4] N. Pryds,[1] and S. Linderoth[1]

[1]*Department of Energy Conversion and Storage, Technical University of Denmark, Risø campus, 4000 Roskilde, Denmark*

[2]*Nano-Science Center, Department of Chemistry, University of Copenhagen, 2100 Copenhagen, Denmark*

[3]*Center for Electron Nanoscopy, Technical University of Denmark, 2800 Lyngby, Denmark*

[4]*Institute of Physics and Beijing National Laboratory for Condensed Matter Physics, Chinese Academy of Sciences, 100190 Beijing, China*

[*] E-mail:yunc@dtu.dk

**Keywords**: Complex oxide, oxide interfaces, two-dimensional electron gases, alumina, strontium titanate

Transition-metal-oxide materials and their interfaces have played a dominant role in ionic-based solid state electrochemical devices for energy conversion and storage, for example, as the heart of solid oxide fuel cells,[1] gas sensors, and logic devices of electroresistive memories.[2] Complementing these applications in solid state ionics, in recent years there has been rapid progress in exploring oxide interfaces for electronics.[3-8] In particular, the progress in atomic scale control of complex oxide film growth has resulted in the discoveries of a quasi-two-dimensional electron gas (q2DEG) at the heterointerface between two band-gap insulators of perovskite $LaAlO_3$ (LAO) and $SrTiO_3$ (STO),[7] and more recently, a 2DEG with extremely high carrier mobilities, exceeding 100,000 $cm^2V^{-1}s^{-1}$ at 2 K, at a epitaxial spinel/perovskite interface between gamma-alumina ($\gamma$-$Al_2O_3$, GAO) and STO (GAO/STO).[8] In reminiscent of the realization of high-mobility 2DEGs at epitaxially grown interfaces made of traditional semiconductors, which has led to a wealth of new physical phenomena as well as new electronic and photonic devices over the past few decades, oxide 2DEGs provide opportunities for a new generation of all-oxide electronic devices.[3-6] On the one hand, like semiconductors, most complex oxides are closely lattice-matched to one another and lead themselves to epitaxial growth. On the other hand, the electrons with partially occupied *d*-orbitals in transitional metal oxides exhibit stronger correlations to other electrons and the lattice. This can give rise to a variety of extraordinary physical phenomena and functionalities at oxide interfaces well beyond those exhibited by conventional semiconductor interfaces, such as 2DEGs with superconductivity or magnetism.[9, 10] Up to date, high mobility oxide 2DEGs are almost exclusively fabricated at temperatures higher than 600 ºC.[3-13] It has become evident that the high-temperature fabrication procedure, firstly, results in strong cation intermixing/diffusion across the interface,[11] which most likely has a deleterious effect on carrier mobilities. For example, the electron mobility of the intensively investigated LAO/STO (deposited at 600-850 ºC) is typically 1000 $cm^2V^{-1}s^{-1}$ at 2 K.[3-13] Similarly, for the high-mobility GAO/STO (deposited at 600 ºC) with film thicker than 3 unit cells (uc), where cation intermixing become unambiguous, the mobility also falls to around 1000 $cm^2V^{-1}s^{-1}$ at 2 K.[8] Secondly, at high temperatures, the oxygen ions in STO can diffuse over many micrometers in minutes.[14] For STO-based heterostructures where oxygen vacancies dominate the interface conduction, such as for the GAO/STO heterostructure,[8] the high temperature process can further level out any nanometer-scale steps in the electron concentration profile along lateral directions, although there can be strong spatial confinement vertically. Even for the LAO/STO system, where the interface conduction is suggested to be dominated by an interface polarity,[7,15] it has also been determined that the length scale of lateral inhomogeneities is as large as 30 $\mu$m when 1 uc thick LAO is epitaxially grown on STO single crystals at high temperatures.[16] Due to the above complexities accompanied with the high-temperature growth of crystalline oxide thin films, it remains extremely challenging to design oxide nanoelectronic devices with electrons confined not only vertically but also laterally, despite recent intensive investigations by different kinds of lithography methods.[17-22]

As room temperature film fabrication process can, without exception, significantly suppress both the cation intermixing and the oxygen bulk diffusion, it is therefore essential to create high quality oxide 2DEGs at room temperature to better understand and utilize their intrinsic properties. Moreover, the room temperature formation of oxide 2DEGs has also the advantage of being compatible with the established lithography of semiconductor microfabrication to pattern oxide interfaces. However, so far, the low temperature growth of complex oxide heterointerfaces, either by pulsed laser deposition (PLD)[23] or by atomic layer deposition (ALD),[24] has resulted in growth of amorphous capping films despite of a metallic interface with electron mobilities around 200 $cm^2V^{-1}s^{-1}$ at 2 K.[23] In this communication, we report the successful room-temperature-creation of oxide 2DEGs at the epitaxial spinel/perovskite GAO/STO heterointerface as illustrated in **Figure 1**a. This novel 2DEG shows electron mobilities exceeding 3000 $cm^2V^{-1}s^{-1}$ at 2 K, which is one order higher than the mobilities of amorphous STO-based heterointerfaces,[23,24] and also approximately three times higher than those typically obtained for crystalline LAO/STO heterostructures.[3-13] The room temperature creation of high mobility oxide 2DEGs, as demonstrated in this paper, could open the door to design oxide nanoelectronic devices with strong carrier confinement in two and three dimensional.

The GAO thin films were grown by PLD, one of the most popular techniques to grow oxide materials. The film growth process was *in-situ* monitored by high pressure reflective high energy electron diffraction (RHEED).[25] Figure 1b shows typical time-dependent RHEED intensity oscillations during the film growth at room temperature. Note that there is a huge transient increase of the RHEED intensity once the film deposition is started, which is followed by a corresponding intensity decrease if the deposition is interrupted. This phenomenon could result from the surface charging due to the insulating nature of the STO single crystal. The most prevalent feature of Fig. 1b is the presence of clear oscillations in RHEED intensity during film growth, which suggests a layer-by-layer two-dimensional film growth mode. Under optimized conditions, high quality GAO films which show streaky RHEED patterns along with clear Kikuchi lines (Supporting Information, Figure S1), are observed up to 32 periods of oscillations. By X-ray reflectivity measurements, it was found that the 32 oscillations corresponded to a film thickness, $t$, of approximately 8 uc. Therefore, the room-temperature layer-by-layer growth of GAO on STO can be controlled on a sub-unit cell scale with one intensity oscillation corresponding to the growth of one quarter GAO unit cell (0.2 nm, the lattice parameter of bulk GAO is 0.7911 nm[26]), similar to the film growth at high temperatures.[8] It should be noted that the RHEED intensity exhibits damping when $t > 8$ uc, which further turns vague as $t > 12.5$ uc. This indicates that the further deposited film evolves into an amorphous state as $t > 12.5$ uc. In other words, there might exist an upper thickness limit for the GAO epitaxial growth at room temperature.

The epitaxial growth of crystalline GAO films on STO is further confirmed by high-resolution X-ray diffraction (Supporting Information, Figure S2) and high-resolution transmission electron microscopy (HRTEM). As shown in Fig.1c, the cross-sectional HRTEM image demonstrates a highly crystalline GAO film epitaxially grown at room temperature, which resembles those samples fabricated at elevated temperatures (600 ºC).[8] Moreover, the spinel/perovskite heterointerface is well-defined with epitaxial relationship of $(001)_{GAO}//(001)_{STO}$ and $<100>_{GAO}//<100>_{STO}$. The lattice spacings measured from the HRTEM images are found to be $a/2=0.392(\pm0.006)$ nm and $c/2=0.401(\pm0.006)$ nm for GAO film and $a=0.393(\pm0.006)$ nm for STO substrate, consistent with the fact that their lattice mismatch is negligible (~1.2%). The film surface morphology measured by atomic force microscopy (AFM) shows an atomic-scale smooth surface with regular terraces, as shown in Fig. 1d. The terraces step is approximately 0.4 nm in height following that of the STO substrate. It is notable that, distinct tiny islands, with a typical diameter of ~100 nm and a maximum height of ~2 nm, are observed along the terrace steps (Fig. 1d). This could explain partially the damping of the RHEED oscillations during the room temperature deposition. Moreover, if these nano-dot islands were crystalline, it would result in spotty RHEED patterns. However, streaky-like RHEED patterns are observed. This suggests that our room-temperature deposited GAO films consist of crystalline flat terraces as well as amorphous nanosize islands along the terrace steps. Additionally, together with the HRTEM results shown in Fig. 1c, the amorphous islands should be present exclusively near the surface region although they were not observed during TEM measurements because of their sparse distribution. In short, high quality GAO films can be epitaxially grown on STO substrates at room temperature in a sub-unit cell layer-by-layer mode. Considering the important role of interlayer mass transport during the layer-by-layer growth of oxide films,[27] the room temperature epitaxial growth of GAO may result from the intrinsic chemical structure of spinel oxides, where the barrier for the cation diffusion can be as low as 0.6 eV.[28] In comparison, the cation migration barrier for perovskite oxides is normally higher than 3.0 eV.[29] Furthermore, it is notable that the $Al_2O_3$ films grown by ALD remain amorphous even at 300 °C.[24] This indicates that the high energy of the incident flux, which is intrinsic to the PLD plume[30] while absent in the other techniques of film growth such as ALD,[24] may also contribute to the room temperature crystallization process observed here. Particularly, it has been reported that the high energy of PLD flux can be rechanneled into enhanced surface diffusion for growing smooth thin films.[31] Note that the room-temperature epitaxial growth of crystalline oxides, such as $CeO_2$,[32,33] has been reported previously by PLD and electron beam evaporation. Additionally, there is an indication that the outward diffusion of oxygen ions from the STO substrate, as discussed in the following, may also play an active role for the epitaxial growth of GAO at room temperature. For instance, the deposition of alumina on a $(LaAlO_3)_{0.3}(SrAl_{0.5}Ta_{0.5}O_3)_{0.7}$ (LSAT) substrate results in the growth of amorphous films. This is because the LSAT substrate can not act as an oxygen source for oxide film growth at room temperature,[23] although it has a similar lattice parameter with GAO. Consequently, the room temperature crystallization of GAO/STO should result from at least three important factors: the small barrier of cation diffusion, the intrinsic high energetic flux of PLD plasma, and the out-ward diffusion of the oxygen from STO substrates.

It has been reported that the GAO/STO interface can become metallic at $t \geq 2$uc when the GAO film is grown at a high temperature of $T_s$=600 °C, [8] despite both GAO and STO are band-gap insulators. Besides the strong dependence of interface conduction on film thickness, the GAO/STO interface created at room-temperature distinguishes from the high-temperature samples by a unique dependence of interfacial conduction on the target-substrate distance, $d$. **Figure 2**a shows a phase diagram for the conduction of GAO/STO grown at different deposition temperatures, $T_s$, for various $t$ and $d$. At $T_s$=700 °C, the heterostructures become conductive once $t > 1$ uc, i.e. the critical thickness for the occurrence of conduction, $t_c \leq 1$uc. These heterostructures show an extremely low sheet resistance in the order of 100 Ω/□ at 300 K and a sheet carrier density, $n_s$, in the order of $10^{15}$ cm$^{-2}$, indicating a mixture of 3D and 2D conduction. For $T_s$=400-600 °C, metallic conduction with vertically spatial confinement, i.e. 2DEG, is observed, which exhibits a constant $t_c$ of 2uc. Note that a 2DEG with unprecedented electron Hall mobilities as large as $1.4 \times 10^5$ cm$^2$V$^{-1}$s$^{-1}$ and a carrier density as high as $3.7 \times 10^{14}$ cm$^{-2}$ at 2 K was obtained at $T_s$=600 °C and $t = 2.5$ uc.[8] Remarkably, decreasing $T_s$ to 300 °C results in a jump of $t_c$ up to 8 uc despite of negligible changes in the sheet resistance. Such conduction persists to $T_s$=100 °C with the same $t_c$ of 8 uc. Unexpectedly, decreasing $T_s$ further to room temperature (20 °C) results in an insulating interface at the normal target-substrate distance ($d$=5.5 cm). In-depth research reveals that the interface conduction of GAO/STO created at room temperature shows strong dependence on $d$. In particular, as shown in Fig. 2b and c, by decreasing $d$ from the conventional 5.5 cm to 4.5 cm, metallic interfaces are achieved in our GAO/STO heterostructures crystallized at room temperature. Similar to other samples grown at higher $T_s$, an interfacial insulator-to-metal transition, which depends critically on the film thickness, is also observed. The heterointerface remains highly insulating at $t<$ 7uc, whereas metallic conduction appears and keeps almost constant once $t>$8 uc (Fig. 2b and c). Note that this threshold for occurrence of metallic conduction ($t_c$=8 uc≈6.4 nm) seems to be an intrinsic factor at $T_s$≤300 ºC, which is four times larger than that for GAO/STO heterointerfaces grown at $T_s$=400-600 ºC (Fig. 2a) and also approximately 4 times larger than those observed in both amorphous and crystalline LAO/STO heterointerfaces.[23,34] Further decreasing $d$ reveals that all the room-temperature-formed interfaces become metallic when $d$≤4.5 cm and $t$≥8 uc.

**Figure 3**a-d shows the representative transport properties of the heterostructures grown at different $d$. For these metallic samples, Hall-effect measurements show a linear dependence of the Hall resistance with respect to magnetic fields (Fig. 3b), where the negative Hall coefficient indicates electron-type charge carriers, *i.e.* n-type. The sheet carrier density, $n_s$, is nearly constant in the whole temperature range of 2-300 K. When decreasing $d$ from 4.5 cm to 2.0 cm, it increases slightly from $8.5\times10^{12}$ cm$^{-2}$ to $2.1\times10^{13}$ cm$^{-2}$ (Fig. 3c, $T$=300 K). The Hall electron mobility, $\mu$, increases upon cooling (Fig. 3d), probably due to the polarization shielding of ionized defect scattering centers driven by the large dielectric constant of STO at low temperatures. Remarkably, a $\mu$≈3200 cm$^2$V$^{-1}$s$^{-1}$ at 2 K is obtained at $d$=2.5 cm (Fig. 3d), which is approximately three times higher than the typical mobility of crystalline LAO/STO formed at high temperatures.[3-13]

Since the Al component in GAO satisfies the chemical criterion for the redox reaction at STO surface,[23,35] i.e. the heat of oxide formation $\Delta H_f^O$<-250 kJ/(mol O) and the work function of the metals, $\varphi$, 3.75 eV<$\varphi$ <5.0 eV, metallic conduction in the room-temperature created GAO/STO heterointerface is therefore expected to result from oxygen vacancies formed as a consequence of interfacial redox reactions. Compared to the high temperature activated redox reaction kinetics, where the relative Fermi levels at the contact play a prominent role,[35] the redox reaction at room temperature is more sensitive to the chemical reactivity of the PLD plasma plume.[30] Provided that the oxidation of the high flux of PLD cation species can be kinetically limited by the availability of sufficient oxygen from the background during film growth, the room-temperature realized 2DEG of the GAO/STO interface at a lower distance should result from a concomitant increased amount of relatively reactive species per square area in addition to a probably enhanced plume front velocity by decreasing $d$.[36] In this vein, a more effective redox reaction is expected at a much lower $d$ as indicated by the increase of $n_s$ with decreasing $d$ (Fig.3c). Additionally, the large threshold thickness ($t_c \approx 8$ uc) for occurrence of 2DEG at room temperature is observed even when $d$ is as low as 2.0 cm. Since this thickness ($\approx 6.4$ nm) is much larger than the typical penetration length ($\leq 1$ nm) of the PLD energetic plasma plume,[23] therefore, the possibility of sputtering induced conduction[37] is ruled out.

To confirm and determine the depth profile of the redox reactions at these room-temperature fabricated GAO/STO heterointerfaces, X-ray photoelectron spectroscopy (XPS) measurements were further performed. **Figure 4**a shows the Ti $2p_{3/2}$ XPS spectra for the GAO/STO interface grown at different distances. No clear $Ti^{3+}$ signal in the $2p_{3/2}$ core-level spectra could be detected in the bare STO substrate, as expected. However, similar to the case of both amorphous and crystalline LAO/STO heterostructures,[23,38] finite amount of $Ti^{3+}$ is already present even in the insulating samples grown at $d$= 5.5 cm. It is noteworthy that decreasing $d$ from 5.5 cm to 4.5 cm, where a sharp insulator-to-metal transition is observed (Fig. 2c, $t$=8 uc), only results in a slight increase in the $Ti^{3+}$ content. This indicates a lower limit of $Ti^{3+}$ content (approximately 7%) to assure interface conduction, the value of which is almost equal to that observed for the amorphous LAO/STO heterointerface.[23] Further decreasing $d$ results in significantly increase of the amount of $Ti^{3+}$ (Fig. 4b), which strongly suggests an enhanced reduction of STO surface as discussed above. For $Ti^{3+}$ depth profiling, angle-resolved XPS measurements were further performed for the sample grown at $d=2.5$ cm. As shown in Fig. 4c, the $Ti^{3+}$ signal indeed shows detectable dependence on the photoelectrons detection angle with respect to the surface normal. An increase of the $Ti^{3+}$ signal with increasing detection angle is determined (Fig. 4d). This suggests that the conduction in our room-temperature formed GAO/STO heterointerface is confined at the interface region. Note that such desired confinement is absent for the other amorphous STO-based heterointerfaces grown at room temperature.[23] If we assume a simple case that the 2DEG layer is stoichiometric, sharp and characterized by a constant fraction of $Ti^{3+}$ per STO unit cell,[8,38] taking into account the attenuation length of photoelectrons, the fitting of the $Ti^{3+}$ content with respect to the detection angle shows that electrons are confined predominantly within a sheet layer of approximately 3.6 nm (Fig. 4d). In short, the XPS results support that interfacial redox reactions account for the conductivity in our GAO/STO heterostructures grown at room temperature as those deposited at high temperatures.[8] It is notable that the 2DEG created at room-temperature shows a non-negligible wider carrier confinement layer compared to those heterostructures grown at higher temperatures ($T_s \geq 300$ ºC) by PLD (Supporting Information, Figure S3), where oxygen ions are more mobile. Further investigation, by polishing the sample surface while continuously measuring the interface conduction, indicates that two different kinds of carrier confinement might exist for the GAO/STO samples grown at different $T_s$ as illustrated in Fig. 4e and f. The carrier confinement in the room-temperature deposited GAO/STO is probably due to the thermodynamic limit of oxygen ions diffusion at room temperature, which results in an almost uniform conducting layer with an extension depth of 3.6 nm. Such assumption is consistent with the high transparency of the sample (inset of Fig. 3b) and the fact that the interface conduction can be removed immediately by slightly polishing the sample surface from the GAO side. Therefore, the room-temperature-created 2DEG at the GAO/STO interface seems analogous to the surface 2DEG in oxygen-deficient STO.[39] Interestingly, the surface 2DEG of STO also exhibits comparable carrier mobilities in the range of 300-2320 cm$^2$V$^{-1}$s$^{-1}$ at 3 K.[39] In contrast, for the GAO/STO deposited at $T_s \geq 400$ °C, the 2DEGs show a much stronger carrier confinement (Supporting Information, Table S1). For example, the 2DEG is strongly confined within a thin layer of 0.9 nm at $T_s$=600 °C.[8] Nevertheless, its conduction remains even after the upmost tens or hundreds of nanometers STO surface is removed by polishing. This indicates that the tail of carriers extends much deeper than the deduced confinement thickness in the GAO/STO formed at higher temperatures. Therefore, a completely different interface confinement might exist at higher $T_s$ as illustrated in Fig. 4f, though the role of high temperature deposition on the carrier mobility needs to be further investigated.

In summary, we have successfully created high-mobility 2DEGs at the spinel/perovskite GAO/STO heterointerface at room temperature by tailoring interface redox reactions. To the best of our knowledge, the obtained mobility of 3200 cm$^2$V$^{-1}$s$^{-1}$ at 2 K is also among one of the highest mobilities for oxide 2DEGs,[3-13,40,41] particularly for patterned complex oxide interfaces. Note that a Hall mobility of ~2800 cm$^2$V$^{-1}$s$^{-1}$ at 2 K is already high enough to observe quantum oscillations in the LAO/STO heterostructure grown at high temperatures (under the condition of a few Tesla for magnetic field and temperature below 1 K).[41] With the compatibility to the established lithography process in semiconductor microfabrication, our finding provides new opportunities to design on-demand oxide nanoelectronic devices.

*Experimental*

The GAO thin films were grown by PLD using a KrF laser ($\lambda$ = 248 nm) with a repetition rate of 1 Hz and laser fluence of 1.5 J cm$^{-2}$. Singly TiO$_2$-terminated (001) STO crystals with a size of 5×5×0.5 mm$^3$ were used as substrates. Commercial single crystalline and ceramic α-Al$_2$O$_3$ were both used as targets, which led to a growth rate of 0.004 nm/s and 0.007 nm/s at room temperature, respectively. During deposition, the oxygen pressure was fixed at a value of $1\times10^{-5}$ mbar for room temperature deposition. For higher temperature deposition, the oxygen pressure was fixed at a value of $1\times10^{-4}$ mbar both during and after deposition. Due to the limit in the beam line setup, *in-situ* RHEED measurements were only performed at $d$=4.5 and 5.5 cm. The film thickness was determined by both RHEED oscillations and X-ray reflectivity measurements. The epitaxial growth of the crystalline film on STO substrates was also confirmed by high-resolution X-ray diffraction measurements (X'Pert Pro) using a Cu Kα X-ray source and high-resolution TEM (HRTEM, FEI Titan 80-300) of cross-sectional samples prepared by focused ion beam (FIB) milling. The film surface morphology was investigated by atomic force microscopy (AFM). The sheet resistance and carrier density of the buried interface were measured using a 4-probe method with ultrasonically wire-bonded aluminum wires as electrodes. Conduction measurements in the Van der Pauw and the Hall-bar geometry were both checked, which gave consistent results. The temperature dependent electrical transport and Hall-effect measurements were performed in the temperature range from 300 K down to 2 K with magnetic fields up to 15 T. The XPS measurements were performed for samples grown at different distances whereas with a fixed film thickness of 8 uc. Before measurements, the thick capping GAO films were firstly etched in 4-M NaOH solution for 3 seconds. This makes the detection of interfacial Ti possible by our XPS instrument, while keeping the interface chemistry largely uninfluenced. For analyzing the Ti $2p_{3/2}$ peaks ($Ti^{4+}$ is at a binding energy of 459.5 eV, whereas the $Ti^{3+}$ is 1.8 eV ± 0.1 eV lower), a Shirley background was subtracted and the spectra were normalized to the total area below the Ti peaks ($[Ti] = [Ti^{4+}] + [Ti^{3+}] = 100\%$).

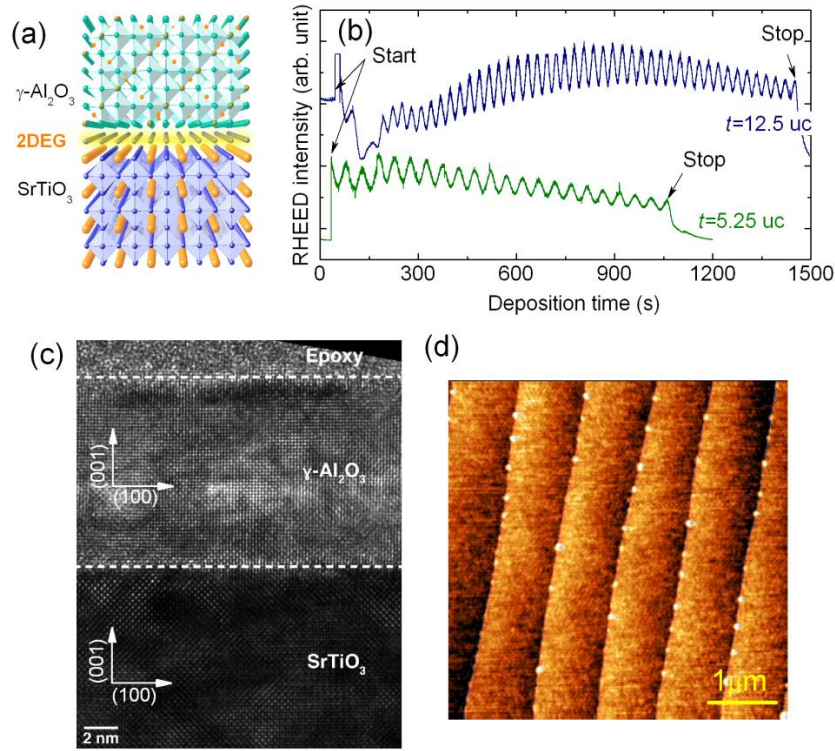

**Figure 1.** Room-temperature layer-by-layer epitaxial growth of γ-Al$_2$O$_3$ **(**GAO**)** films on SrTiO$_3$ (STO). (a) Sketch of the 2DEG at the spinel-perovskite GAO/STO interface; (b) Time-dependent RHEED intensity for the film growth by ablating from a ceramic (blue) or a single crystalline (green) α-Al$_2$O$_3$ target; (c) A cross-section HRTEM image of the epitaxial GAO/STO interface formed at room-temperature; (d) A representative surface morphology of the GAO film with tiny islands, approximately 100 nm in diameter and 2 nm in height, along terrace steps. The terrace height is 0.4 nm.

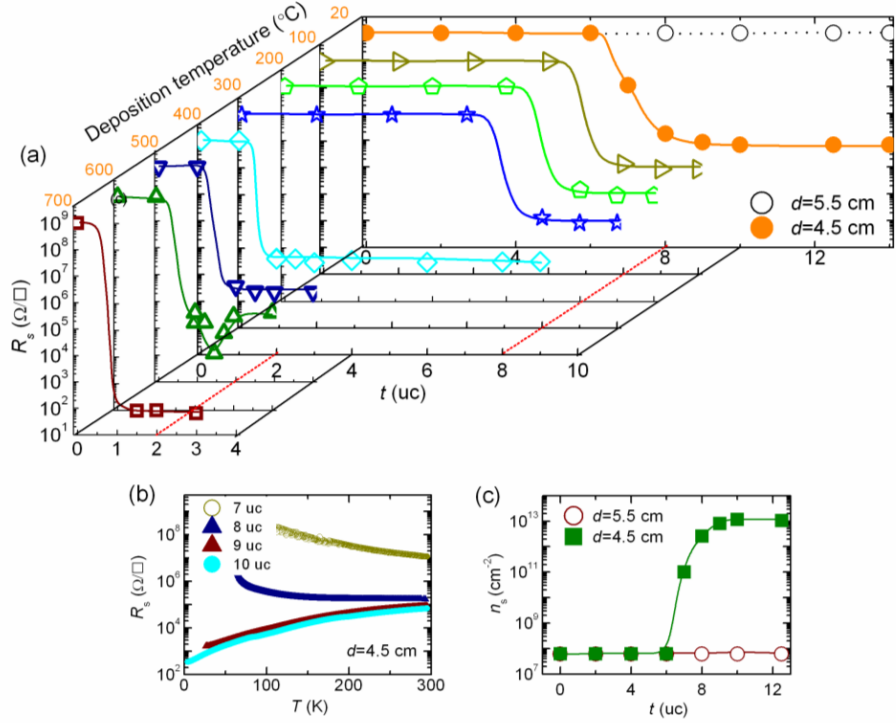

**Figure 2.** (a) Phase diagram for the conduction of GAO/STO interfaces grown at different temperatures. The room-temperature-formed interface shows a unique dependence of conduction on target-substrate distance, $d$, (open symbol for $d=5.5$ cm, and solid symbol for $d=4.5$ cm); (b) Temperature dependence of sheet resistance, $R_s$, at different film thicknesses, $t$, for $d=4.5$ cm; (c) Thickness dependence of the carrier density, $n_s$, measured at 300 K. High-mobility 2DEGs are obtained once $t$ is above 8 uc at $d=4.5$ cm for room temperature deposition.

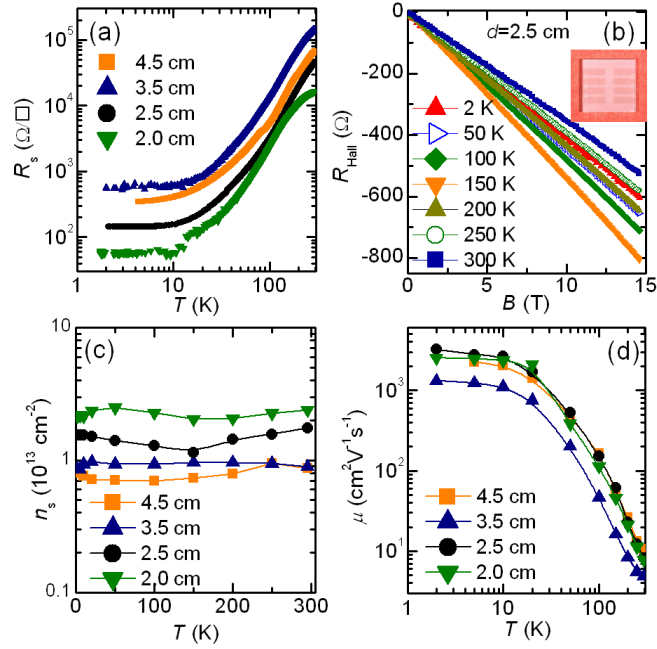

**Figure 3.** Transport properties of the GAO/STO interface grown at different target-substrate distances. (a) Temperature dependence of sheet resistance, $R_s$; (b) Hall resistance $R_{Hall}$ versus magnetic field at different temperatures for $d$=2.5 cm, A linear Hall effect is observed in the whole temperature range from 300 K to 2 K; (c) and (d) Temperature dependent carrier density, $n_s$, and electron Hall mobility, $\mu$, respectively. A remarkable $\mu$=3200 cm$^2$V$^{-1}$s$^{-1}$ at 2 K is obtained at $d$=2.5 cm. Inset of (b) shows a photo image of the sample with a Hall-bar on surface.

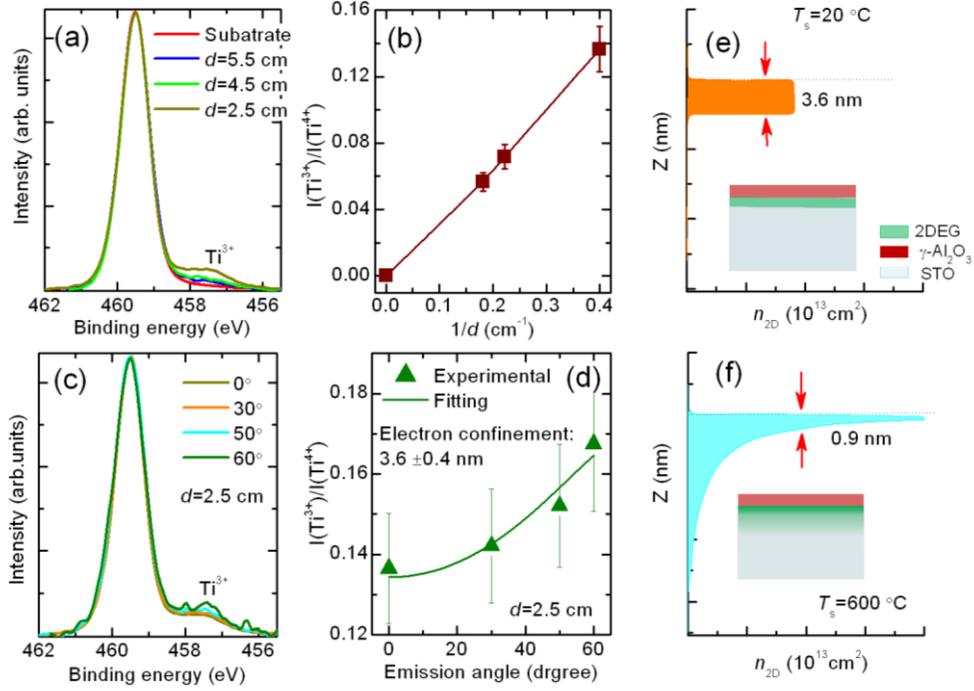

**Figure 4.** (a) The Ti $2p_{3/2}$ XPS spectra for the GAO/STO interface grown at different distances; (b) the ratio of $Ti^{3+}$ to $Ti^{4+}$ signal, $I(Ti^{3+})/I(Ti^{4+})$, versus $1/d$. The increase of $Ti^{3+}$ content with decreasing $d$ suggests an enhanced chemical redox reaction at the interface; (c) The Ti $2p_{3/2}$ XPS spectra at various emission angles for the $d=2.5$ cm sample; (d) The angle dependence of $I(Ti^{3+})/I(Ti^{4+})$, indicates a confinement of the conduction layer within 3.6 nm; (e) and (f) Illustration of the different confinement of 2DEGs in GAO/STO obtained at room temperature ($T_s=20$ ºC) and $T_s=600$ ºC, respectively.